%% file: main.tex
\documentclass[aps, prd, a4paper, 10pt, showpacs, superscriptaddress, floatfix, nofootinbib, reprint]{revtex4-2}
\usepackage{color}
\usepackage{xcolor}
\usepackage{amsmath,amssymb,graphicx}
\usepackage{wasysym}
\usepackage{bm,lipsum}
\usepackage{times}
\usepackage{multirow}
\usepackage{microtype}
\usepackage{comment}
\usepackage[utf8]{inputenc}
\usepackage{booktabs}
\usepackage{subfigure}
\usepackage[normalem]{ulem}
\usepackage[varg]{txfonts}
\usepackage{bm,graphics,graphicx,epsfig,color,ulem}
\usepackage{array}
\usepackage{tabularx}
\usepackage[colorlinks, pdfborder={0 0 0}]{hyperref}
\usepackage{bm}

\definecolor{LinkColor}{rgb}{0.75, 0, 0}
\definecolor{CiteColor}{rgb}{0, 0.5, 0.5}
\definecolor{UrlColor}{rgb}{0, 0, 0.75}
\definecolor{maroon}{rgb}{0.5, 0.0, 0.0}
\definecolor{mint}{rgb}{0.24, 0.71, 0.54}
\definecolor{violet}{rgb}{0.62, 0.0, 1.0}
\hypersetup{linkcolor=LinkColor}
\hypersetup{citecolor=CiteColor}
\hypersetup{urlcolor=UrlColor}
\allowdisplaybreaks
\newcolumntype{C}{>{\arraybackslash}m{6cm}}
\newcolumntype{?}{!{\vrule width 1pt}}

\usepackage{lineno}

\begin{document}

\title{Performance of iDQ ahead of LIGO, Virgo, and KAGRA’s fourth observing run}
\input{authors}

\begin{abstract}
    The gravitational wave detectors used by the LIGO Scientific Collaboration, and the Virgo Collaboration are incredibly sensitive instruments which frequently detect non-stationary, non-Gaussian noise transients.
    iDQ is a statistical inference framework which leverages the use of auxiliary degrees of freedom monitored in the detectors to identify such transients.
    In this work, we describe the improvements to the iDQ pipeline made between the third and fourth observing run of the LIGO-Virgo-KAGRA (LVK) collaboration, and show the performance of these changes.
    We find that iDQ detects a total of 39,398 of the known 100,512 glitches identified by Omicron over the course of the second half of the third observing run.
    We construct a measure of the probability a glitch is present in the strain data of a given detector by combining information from iDQ and Omicron as well as extend the output of iDQ in a novel method which finds correlations between known glitch classifications, and auxiliary channels.
    We identify several channels over the course of O3b which frequently record instances of Scattered Light, Whistle, and Blip glitches and discuss use cases for this method in active observing runs.
    
\end{abstract}

\maketitle

\footnotetext[2]{\href{rngeorge@utexas.edu}{rngeorge@utexas.edu}}

\section{Introduction}
The detection of 90 gravitational wave candidates \cite{gwtc-3:2021vkt} by the
LIGO Scientific Collaboration and the Virgo Collaboration has been made possible via gravitational wave detectors, Advanced LIGO  \cite{Aasi:2015aLIGO} and Advanced Virgo \cite{Acernese:2015Virgo}.
The detectors are modified large-scale Michaelson interferometers whose main output, called strain data, is the difference in distance traveled by laser light between its two arms.
As gravitational waves pass through the instrument, these paths change a minute amount, resulting in a recorded signal in the strain data measured to lengths as small as $10^{-19}$ meters (when measuring within the sensitive frequency band).
While this remarkable sensitivity allows for the detection of gravitational waves, it also allows for the easy detection of transient noise sources arising from the environment or the instrument itself.
Such non-stationary noise sources are commonly referred to as glitches.

Glitches can arise from a variety of both known and unknown sources \cite{Nuttall:2018xhi, Cabero:2019orq,LIGOScientific:2019hgc,LIGO:2021ppb} and pose a challenge to the accurate detection and analysis of gravitational wave signals \cite{LIGOScientific:2016gtq,Canton:2013joa,LIGOScientific:2017tza, Powell:2018csz}.
Some glitches can manifest in the strain data in a similar manner to genuine gravitational wave signals, and can possibly be mistaken for them if present simultaneously across multiple detectors.
Glitches can additionally overlap with true signals thereby obfuscating them, and making parameter estimation of the true signal difficult.
Such was the case with the detection of the binary neutron star merger signal GW170817 where there was a large glitch overlapping with the Livingston detector data causing some initial concerns with the data quality \cite{LSC:gw180817, Pankow:2018qpo}.
Therefore identifying and characterizing glitches is a key part of increasing the overall sensitivity of the entire detection system.

While the detection of gravitational waves is possible using the strain data alone, there are also thousands of supplementary data outputs produced by the detectors.
These supplementary outputs, called auxiliary channels, record additional degrees of freedom in the detector apart from the strain data and act as monitors on everything from mirror deformation, to environmental recordings \cite{Matichard:2015eva, GraefRollins:2016xhy, aLIGO:2016pgl, Effler:2014zpa, LIGOScientific:2019hgc}.
The presence of a gravitational wave signal in the main channel can be seen in a subset of these auxiliary channels. Such channels are not ideal for use in identification of glitches as signals there could be genuine gravitational waves signals.
Safe auxiliary channels then are defined those which are insensitive to gravitational waves, and therefore any signal present in these channels is by definition a glitch.
Glitches witnessed by these safe auxiliary channels may also appear in the strain channel.
If only monitoring the strain channel in such a case, information is lost that could easily identify the present signal as a glitch.
iDQ is a statistical inference framework which uses safe auxiliary channels to identify these cases and make statistical statements about the presence of glitches in the strain channel based solely on activity in the auxiliary channels \cite{Essick:2020qpo, Godwin:2020isc}.

iDQ is trained on activity in safe auxiliary channels labeled by the presence of glitches in the strain data to identify correlations between the two.
If an auxiliary channel is identified to be strongly correlated with the strain data, then any new activity in that auxiliary channel can be used to predict if a signal in the strain data is of terrestrial origin.
Additionally, by monitoring only the safe auxiliary channels, iDQ can safely identify glitches without also flagging real gravitational waves making it ideal for incorporation into gravitational wave detection pipelines.
The output of the iDQ analysis then consists of two probability statements that indicate the likelihood that the gravitational wave data is contaminated by a glitch monitored by one of these auxiliary channels.

There are thousands of these safe auxiliary channels sampled at high rates available for analysis by iDQ.
In order to reduce the computing cost and latency of analyzing so many channels at such high rates, iDQ relies on two sources for the extraction and downsampling of relevant information.
In low-latency operation, the Stream-based
Noise Acquisition and eXtraction pipeline (SNAX) \cite{Godwin:2020isc}, is implemented for this purpose, but in high latency Omicron \cite{Robinet:2020lbf, McIver:2015pms} is used and it is the latter which will be discussed in this work.
Omicron reports on the presence of excess power, measured by the signal-to-noise ratio (SNR), in both strain and auxiliary channel data via the Q transform, a wavelet decomposition which uses sinusoid signals modulated by a Gaussian amplitude and is parameterized by frequency, amplitude, and a quality factor (Q).
Times which are noted to have an estimated SNR greater than 5.5 are then passed on to iDQ for analysis.
High SNR times from the strain channel are used to label glitches in iDQ training sets, while all times from auxiliary channels are used to find correlations.

Correlations identified by iDQ can then be further classified by the type of glitches present in the correlation.
Glitches are separated into classes based on how they appear in the strain data and those which have the same morphology as gravitational wave signals, for example, are of particular interest and are targeted for further study.
Automating the classification of glitches is what motivated the development of GravitySpy \cite{Glanzer:2022avx, Bahaadini:2018git, Soni:2021cjy, Mukund:2016thr, Wu:2024tpr} which can automatically classify any time of interest using a convolutional neural network (CNN) and a training set of known morphologies.
In this work, the correlations iDQ finds between auxiliary channels and strain data are extended using the classifications assigned by GravitySpy to find relationships between auxiliary channels and glitch classes.
These relationships then reveal which glitch classes appear most frequently in which auxiliary channels.
If an auxiliary channel frequently witnesses a particular glitch class, then the detector system the auxiliary channel monitors can be investigated as a possible source of that glitch class.

In this work, Section II provides a background of the iDQ framework, the Omicron package it relies on, and the GravitySpy package which provides classifications. Section III reviews changes to the iDQ pipeline from the LSC's third observing run to the beginning of the fourth run, as well as the performance of these changes.
Finally, in Section IV we quantify iDQ's current performance, including a new measure of glitch probability, and report on a new method using iDQ to track glitch types back to their possible origins in the detectors.

\section{Background}
\label{sec:background}
\subsection{Omicron}
\label{sec: omicron}
Omicron is designed to detect and characterize transient signals through the use of the Q transform, which decomposes the detectors' time-series data into a time-frequency basis.
The Omicron implementation of the Q transform relies on the tiling of a signal's time-frequency space where one tile is defined by the projection of the signal onto a Bisquare windowed sinusoid basis with a given central time, central frequency, and quality factor, Q \cite{Robinet:2020lbf}.
The distribution of tiles used by Omicron in a single Q plane is defined by an acceptable energy loss due to mismatch between the tiles, and this strategy leads to sets of tiles defined logarithmically in central frequency and Q, and linearly in central time.

The excess energy of any given tile is then used as an estimation of the SNR, $\rho$ and is given as \cite{McIver:2015pms}:
\begin{equation}
    \rho^2 = \frac{|X(\tau, f_c, Q)|^2}{<X(Q)^2>}-2
\end{equation}
where $X(\tau, f_c, Q)$ is a single tile, $<X(Q)^2>$ is the mean expected energy of all tiles in a given Q plane, and 2 is the result expected from white noise.
In order to form triggers, any two tiles with positive excess energy which have less than 0.1 seconds between their central times are considered to be identifications of the same event and are clustered.
After a cluster is formed, the SNR, central time, and central frequency of the tile in the cluster with the highest SNR are assigned as the event's parameters.
Similarly, the start and stop time of the event are taken as the earliest and latest central time of tiles in the cluster \cite{McIver:2015pms}.
This clustering results in events with non-uniform duration from as little as 0.1 seconds and up to 10 seconds, with the bulk of the distribution around at $\mathcal{O}$(0.1) seconds.

Any event with an SNR of at least 5.5 in the strain channel is then recorded in the Omicron database along with its relevant parameters.
High SNR events from the strain channel could indicate the presence of a glitch, but could just as easily identify a real gravitational wave signal.
For the purpose of this paper, events with an SNR greater than 10 in the strain channel which do not correspond to known gravitational wave signals are treated as the identification of a glitch.
Events identified in safe auxiliary channels, meanwhile, are used as inputs to the iDQ analysis.

\subsection{GravitySpy}
While other detector characterization pipelines like Omicron identify glitches in the strain data, GravitySpy contributes classifications to those times.
There are variety of glitch classes defined by detector characterization experts based on their morphology in the time frequency space of the strain channel \cite{Glanzer:2022avx}.
GravitySpy uses a combination of human volunteers and a CNN machine learning algorithm \cite{Bahaadini:2018git,Zevin:2016qwy,Soni:2021cjy, Wu:2024tpr} to classify any time of interest based on spectograms of the strain data with four different durations.
Four durations are used in order to expose morphologies which are present at varying timescales.
The output of GravitySpy tells us at a given time what is the probability, or confidence, of each glitch class considered.
The confidence across classes is not required to sum to one, and instead each class is assigned a value between 0 and 1 individually.
For example, GravitySpy could be confident that there was a scattering-like glitch present, but be unable to distinguish whether it was Fast Scattering or Scattered Light.
In this case, you could expect high confidence values in those two classes and lower confidence values across all others.

The training set for GravitySpy has evolved with time as more glitches are discovered, and more morphological classes are defined.
As of LIGO's third observing run, the training data consisted of 9631 labeled glitch samples across 23 morphologies \cite{Glanzer:2022avx}.
There has since been an update to the GravitySpy model for LIGO's fourth observing run \cite{Wu:2024tpr}, but this paper focuses on data from the third and therefore uses classification from the GravitySpy model available during that time.

During active LIGO observing runs, GravitySpy classification is triggered on new Omicron event times uploaded to the database with SNR of 7.5 or greater.
This allows for medium-latency identification and classification of glitches which can then be used by detector engineers to inform detector maintenance.
However, Omicron experts frequently add new events to the database which were not identified in low-latency.
These additional times are not always assigned classifications by GravitySpy, and therefore in this work, we consider an additional category of "Unclassified" to represent these times.

\subsection{iDQ}
We will give a brief summary of the iDQ framework leading into the LSC's third observing run, further description can be found in \cite{Essick:2020qpo}. 
iDQ runs in two modes -- streaming and batch. We summarize the batch, or high-latency, implementation here as it is what is used to measure the performance of the pipeline in the later sections of this work.

iDQ begins by taking in events produced by Omicron on O($10^{3}$) auxiliary channels.
As discussed in section \ref{sec: omicron} triggers are equivalent to tabular data on transients in these channels and contain information on the signal-to-noise-ratio (SNR), frequency, central time, etc of these transients.
The events for each channel reported by Omicron are then downsampled further by iDQ into feature vectors.
A feature vector represents the maximum SNR event reported by Omicron in any one second window.

These vectors are then labeled as glitch or clean based on the SNR value reported by Omicron on the strain data at the same time.
If the SNR of the strain channel is greater than 10, then the transient is considered a glitch. 
If the strain channel SNR is less than 5.5, then its considered clean.
Any feature that falls in between those two thresholds is neither clean nor a glitch and is not used in training data.

To construct training datasets, the entire time of interest is first divided into segments.
The majority of these time segments are used for training, and one is reserved for evaluation, as described in detail in Section IV of \cite{Essick:2020qpo}.
The times in training segments are then used to construct the training datasets iDQ needs for its classifiers.
For training, all the times labeled as glitch, and a random selection of clean times at least one second away from a labeled glitch are used.
The additional one second window for clean sampling enforces that the times in the glitch and clean datasets are uncorrelated as most glitches are shorter than one second in duration. 
This limits the training dataset based on the Omicron SNR threshold in strain, but it is ultimately the auxiliary channel features at the times of interest, and not strain information that iDQ is trained on.

iDQ uses these datasets to then train the classifier(s) chosen.
The only limit to the choice of classifier is that it must map the high-dimensional input feature vector space into a single rank value between 0 and 1. 
This mapping, called a model, is unique to the classifier and to the time it was trained on. 
For batch production during O3, the classifier OVL was used.

The OVL classifier is described in \cite{Essick:2013vga}, but can be summarized by understanding that models produced by OVL consist of an Ordered Veto List, a list of times during which the activity in the auxiliary channels indicates the presence of noise in the strain channel. To form this model, OVL creates a list of possible veto configurations based on auxiliary channel, threshold, and time window and then evaluates them based on a chosen metric.
OVL previously supported the choice of one of three different metrics for this process: efficiency-deadtime ratio, Poisson significance, or use percentage, although as we describe in section \ref{subsection: change of rank}, additional options have been added.
The use percentage is the fraction of auxiliary channel glitches which can be associated with a strain channel glitch where a glitch in the auxiliary channel is defined by having an amplitude above the threshold set by the veto.
The Poisson significance is the probability of observing as many or more coincidences between two series of random events than actually observed between the auxiliary channel and strain as described in detail in \cite{Smith:2011an}.
The efficiency-deadtime ratio is given by the efficiency of the veto over the deadtime introduced by the application of the veto.
In other words, the fraction of total glitches in strain removed over the fractional livetime removed by applying the veto.

iDQ requires that these metrics produce ranks that fall in the space of [0,1) rank space in order to be comparable between one another.
If only used individually, the choice of [0,1) is an arbitrary one and what is more important is the ordering.
However, as will be discussed in section \ref{sec:improvements}, it can be desirable to combine information across several metrics in order to give preference to vetoes with a certain combination of properties.
Therefore, a simple scaling factor was applied to their values with a map from the metric space of [0, inf) to rank space of [0, 1) as show below:
\begin{equation}
    Rank_m = \frac{s_m * x_m}{1 + s_m * x_m}
    \label{eq:old_rank}
\end{equation}
where $s_m$ is the scaling factor for the metric and $x_m$ was the value of the metric itself.

After the initial rank evaluation, the vetoes are ordered from highest to lowest, and then the rank is re-calculated applying the highest veto first, and ending in the lowest.
Any veto falling under a threshold for that metric is removed, and the process is repeated.
In this way, OVL produces a final Ordered Vetoed List for any given training time, thereby making a model.
The model for OVL is then applied to a time of interest by first removing any veto configuration which doesn't apply.
Then, the rank of the highest ranked veto from the resultant list is applied as the rank of the time of interest.

The rank values alone, however, do not have any physical meaning and it is only the ordering that truly matters.
Therefore, ranks must be transformed to log-likelihood and false alarm probabilities through calibration. The calibration requires the rate of clean and glitch samples to be determined, and a PDF generated.
Through O3, the glitch and clean distributions for this calibration were populated by the glitch and clean samples from the training datasets, although this has changed recently as described later in section \ref{subsection: all data for background}.
The rates of clean samples and glitch samples can then be calculated directly from these distributions as described in \cite{Essick:2020qpo}, and a Gaussian kernal-density-estimate (KDE) can be applied to the distributions to obtain a posterior-density-function (PDF).
These rates, along with the PDF from the KDE, create the calibration maps needed to convert any given rank to a log-likelihood and false alarm probability as described in detail in \cite{Essick:2020qpo}.

In summary, for any time of interest, a model made by OVL and trained using Omicron labels on the strain is applied to the full feature vector set.
The result is a single rank.
The rank is then transformed to log-likelihood and false alarm probability statistics via a calibration map generated by sampling the clean and glitch distributions of the training datasets and applying a Gaussian KDE.
This process can then be repeated for any number of times, thereby creating a full timeseries.

\section{Pipeline Improvements}
\label{sec:improvements}
The methods described in \ref{sec:background} were applied uniformly through O3.
However, between the end of O3 and the beginning of O4 in May 2023, several changes were implemented into the iDQ analysis to improve calibration, and the dynamic range of its outputs.
These changes were applied to batch offline re-analysis of O3 data in order to prepare detection pipelines for the fourth observing run (O4) and then applied to both the batch and streaming iDQ analyses during the first half of O4.
In the following three sections, we describe in detail the changes and the motivations behind them.
In \ref{sec:performance}, we report on the performance of these changes and show that they lead to the desired effects.

\subsection{Change of Rank}
\label{subsection: change of rank}
The first of these changes was to the calculation of the rank assigned to times of interest by OVL.

The metrics of use percentage, Poisson significance, and efficiency-deadtime ratio individually are useful, but previously it was difficult to compare results across them as they did not behave similarly across the rank space.
A similar behavior of increasing metric leading to increasing rank with support throughout the [0, 1) rank range was needed for each metric.
To get this behavior, the scaling factors described in Equation \ref{eq:old_rank} for the efficiency-deadtime ratio and Poisson significance were altered but otherwise the rank is calculated identically.
Meanwhile, the use percentage to rank map was changed entirely to have no scaling factor or map at all.
That is, the rank is exactly the use percentage.

With this change, we then offer the option to combine ranks from multiple sources at once to obtain weighted ranks.
This is done via a weighted average of the three individual metrics as shown below:
\begin{equation}
    \frac{\sum{r_m * w_m}}{\sum{w_m}}
    \label{eq:new_rank}
\end{equation}
where $w_m$ is a configurable parameter and is weight for a given metric, and $r_m$ is the value of the rank from a given metric.
In a typical offline analysis, the veto efficiency and use percentage of the veto are used with a weight of one third and two thirds respectively.
This configuration down-ranks vetoes with high efficiency but poor use percentage.
Vetoes which follow this trend flag auxiliary channels with rare departures above the threshold but in the process veto large periods of quiet time in the strain channel, increasing the likelihood that iDQ misidentifies a genuine event as being of instrumental origin.
By prioritizing the use percentage metric, these veto configurations are suppressed in favor of those which are less likely to misidentify genuine events in the strain channel as noise.

\subsection{Use of all data for background collection}
\label{subsection: all data for background}
In order to calibrate the rank from iDQ's classifiers into statistical information such as the false alarm probability and log-likelihood, a model of the underlying distribution of clean and glitch samples is generated as described in \ref{sec:background} using a Gaussian KDE.
However, the KDE can only be as accurate as the underlying distribution it relies on.

As described previously, when using the binning and segment scheme developed for the offline batch mode of iDQ, the clean distribution used in calibration is populated only with times from the training datasets.
This results in populating the clean distribution only with times at most one half of a segment's width away from the time of interest, or typically $\mathcal{O}$(3) days.
When there are few segments for a wide time range, or the segment width is large, the clean distribution used for calibration can be very different from the true distribution which the time of interest resides in.
This is a natural result of the time-evolving nature of the detectors as the noise background one day can vary significantly from the noise background the next.

This results in the output timeseries jumping between segment boundaries as the calibration between those segment boundaries reflects the change in the underlying noise distribution.
By allowing sampling of the clean distribution within the time segment of the time of interest in addition to sampling outside of it, the resulting clean distribution more accurately reflects the local distribution, and the output timeseries becomes more seamless between segment boundaries.

\subsection{Bounding of KDE Bandwidth}
\label{subsection: bounding}
As mentioned in section \ref{sec:background}, iDQ applies a Gaussian KDE to the discretely sampled clean and glitch histograms to create smooth posterior density functions.
These distributions are then converted to cumulative distributions in order to calculate the false alarm probability (FAP) and log-likelihood.
The accuracy of this KDE then has a direct impact on the timeseries output of iDQ.

\begin{figure}[htb]
    \centering
    \includegraphics[width=0.48\textwidth]{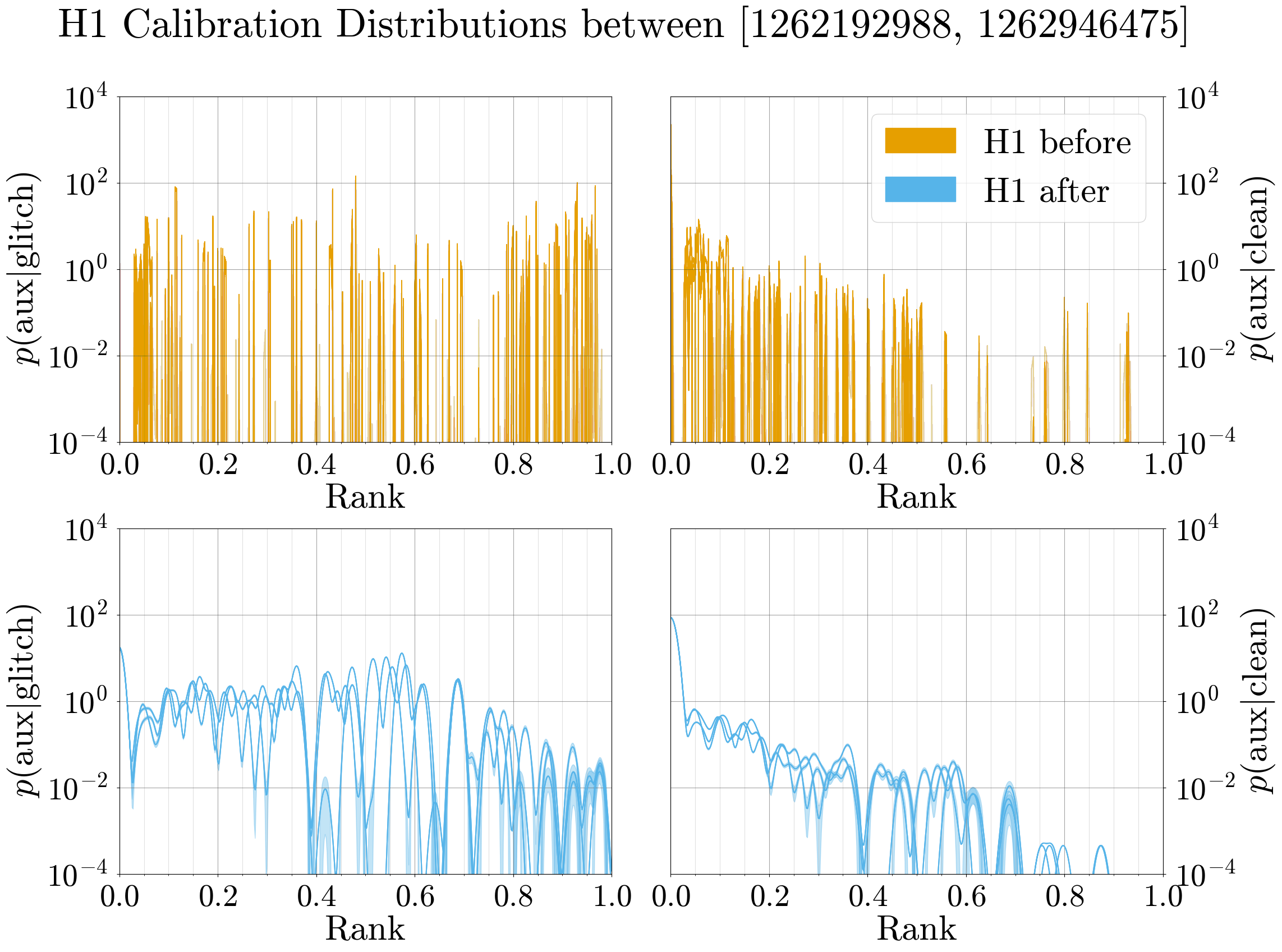}
    \caption{In the top row, the calibration distribution is shown before the KDE is applied. In the bottom row, the KDE of the glitch distribution (left) and clean distribution (right) of samples collected over about two weeks of O3b time after the most recent changes were implemented. Note how the KDE is significantly smoother than before and provides a wide range of support across the rank space. This results in the output statisitics based on this KDE being more evenly scaled across the entire rank space.}
    \label{fig:kde_bound}
\end{figure}

\begin{figure}[htb]
    \centering
    \includegraphics[width=0.48\textwidth]{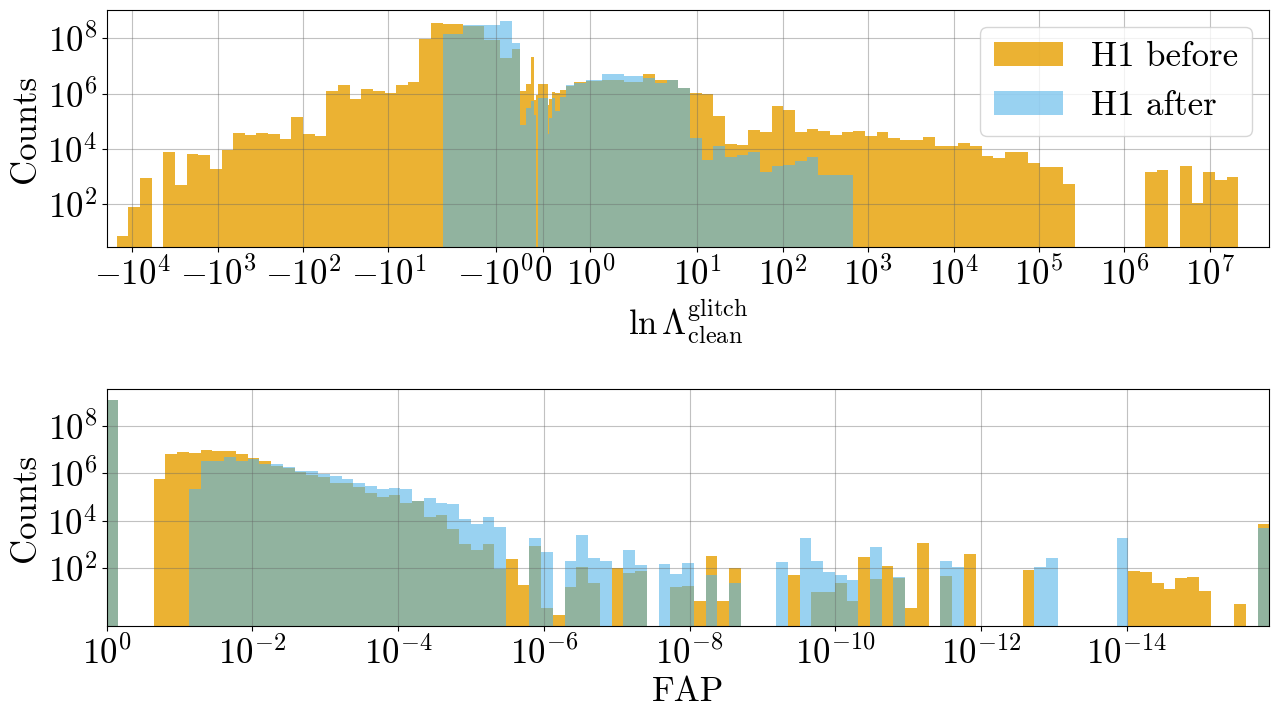}
    \caption{Histograms showing the count of samples with log likelihood ratio between the glitch and clean models (top) and false alarm probability (bottom) observed by iDQ at Hanford during O3b both before code changes were implemented (yellow) and after (blue). Notably, there is a vast improvement in the dynamic range of the log-likelihood distribution after the code changes as desired.
}
    \label{fig:histogram_new}
\end{figure}

Previously, the discrete nature of the clean and glitch distributions caused the automatic bandwidth optimization of the KDE to rail to extremely narrow Gaussians as shown in the top half of Figure \ref{fig:kde_bound}.
This left large regions of the rank space without proper support from the KDE, especially as rank goes to one.
This caused scaling of the log-likelihood and FAP to be uneven, and the log-likelihood in particular to have a large dynamic range.
Bounding the lower end of the bandwidth range forces support in those portions of rank-space without many samples in its underlying histogram as seen in the bottom half of Figure \ref{fig:kde_bound}.
This results in a more evenly scaled output statistics, with a smaller dynamic range.

Improvements in the distribution and bounding of the timeseries can be seen in Figure \ref{fig:histogram_new}.
Here, we only show the results for Hanford because the Livingston timeseries show similar behavior changes.
The top half of this plot shows the histogram of the log likelihood ratio between the glitch and clean models with the original timeseries shown in orange and the improved version in blue.
The dynamic range of the log-likelihood ratio has been severely reduced from thirteen orders of magnitude to just six with the updates.

\subsection{Performance of Pipeline Changes}
\label{sec:performance}
In order to show the improvements made in the pipeline, we compared results from iDQ's analysis of O3b, which spans from November 2019 to March 2020.
We compare the results from the original offline analysis done concurrently with O3b observations and from a re-analysis using the newly updated pipeline code.
For simplicity we will refer to the different code versions as being 'before' and 'after' respectively.

\begin{figure}[htb]
    \centering
    \includegraphics[width=0.48\textwidth]{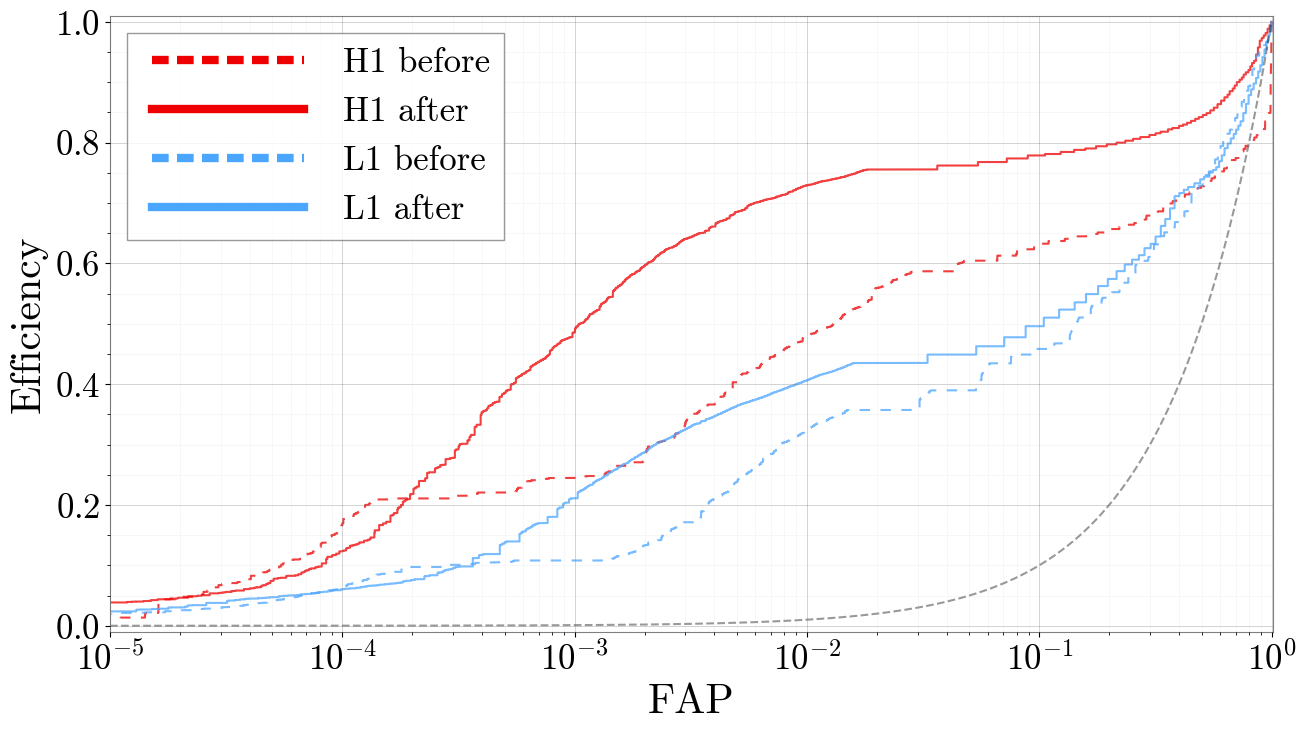}
    \caption{Receiver operating characteristic curve of iDQ at the Hanford (red) and Livingston (blue) gravitational wave detectors over the time period of O3b. The dashed lines represent the original results from iDQ and the solid lines show the re-analysed results after code improvements. A dashed grey line was added to represent an uniformed classifier. The analyses at both detectors show improvement in the important mid-FAP range where the bulk of iDQ's distinguishing power.
}
    \label{fig:roc_new}
\end{figure}

Figure \ref{fig:roc_new}, shows the false alarm probability (FAP) plotted against the efficiency of the iDQ pipeline at Hanford (red) and Livingston (blue) with original code (dashed) and the updated version (solid).
In these receiver operating characteristic (ROC) curves, we can already see a stark contrast between the two code versions -- particularly in the performance at the Hanford detector.
Both at Hanford and Livingston, there is a doubling of the efficiency of the pipeline at a false alarm probability (FAP) of $10^{-3}$ after the updates and a general improvement across the range of FAP $10^{-3}$ to $10^{-1}$.
At larger FAP, the model for Hanford previously did worse than an uninformed one, or a model built on random chance, but after the updates this is no longer the case.
However, the Livingston performance is slightly worse at high FAP than before.
Generally, this slight decrease in performance at high FAP is seen as a more than fair trade-off for the wide improvement in the middle range.
At high FAP, iDQ loses its distinguishing power as there is a gap in the rank output of OVL between 0, and the lowest ranked veto configuration as can be seen in Figure \ref{fig:histogram_new}.
This means that iDQ already does not have distinguishing power in the FAP range, so a small loss of sensitivity there is not a large loss to the power of the analysis.
Instead, its the middle ranges of the FAP where we see the most improvements which are the most crucial to the distinguishing power of the analysis.

\section{Use of iDQ}
iDQ's main purpose is to identify glitches apparent in the strain data by monitoring the auxiliary channels of the detectors.
Omicron strives for the same end goal, but via excess power in the strain channel itself.
In this section, we use the glitch times flagged by Omicron as a benchmark for comparison for the output of iDQ and then analyze the results broken down by glitch class and auxiliary witness channel.

In Section A, we discuss how Omicron and iDQ identify glitches, how we construct coincident events between the analyses, and how the results break down by glitch class. In Section B, we show how iDQ can provide additional auxiliary channel information about identified glitches, and possibly identify physcial sources of glitch classes.

\subsection{Glitch Presence Identification}
\label{sec: glitch_identification}
\subsubsection{Methods}
\label{sec: glitch_identification_methods}

\begin{figure*}[htb]
    \centering
    \includegraphics[width=2\columnwidth]{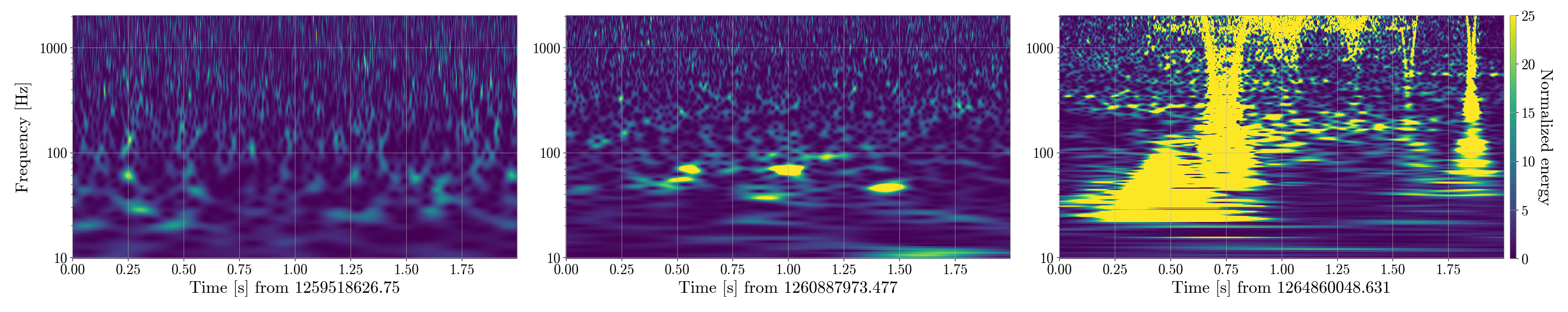}
    \caption{Qscans of three times identified by both Omicron, and iDQ as being a possible glitch, but which was classified by GravitySpy as the category "No Glitch".
    The left-most plot shows a correct classification of a No Glitch time while the middle and right show incorrectly classified times. The leftmost figure shows excess power, but fairly evenly distributed with no clear concentration -- a true No Glitch classification. The middle shows a clear time of excess power in a distinct shape, but not in a morphology that matches any other glitch class. The right most plot shows a time which is clearly an Extremely Loud glitch, but was just mis-classified by the model.}
    \label{fig:no_glitch_qscan}
\end{figure*}

In this work, we compare iDQ's performance against that of Omicron at identifying glitches broken down by glitch type as classified by GravitySpy.
In order to define a glitch as identified by iDQ, we first apply a threshold on the log-likelihood ratio.
In this work, we've chosen two thresholds to examine -- two and five.
We have chosen to analyze both of these because we found certain glitch classes ring up frequently in the 2 to 5 range, but the lower confidence threshold additionally results in more false alarms.
The nature of veto application causes the output timeseries of iDQ to be step-wise with steps at most the width of the largest veto window, but typically less than a second.
Therefore, after application of the threshold, we cluster any points with identical adjacent neighbors by keeping only the central point in time from series of identical points.
This is equivalent to identifying the center of any veto window as the central time of the event.
We then further cluster these points by keeping the maximum log-likelihood point in a clustering window, $w_{iDQ}$, of one second, or a half second on either side.
Using this large clustering window allows us to assume that any two event times identified by iDQ are not caused by the same glitch, and are uncorrelated.

We then compare these events identified by iDQ to all times identified by Omicron in the strain channel as having an SNR of ten or greater.
The threshold of ten on Omicron SNR is chosen to match the SNR threshold used in iDQ training datasets, and we calculate the Omicron glitch rate, $\sigma_{omic}$ as the number of events crossing this threshold by the observing time.
Any iDQ event within a coincidence window, $w_{coinc}$, of a half second on either side of a glitch identified by Omicron we assume is an identification of the same glitch and call this event coincident between the analyses.
We can then count the number of these events for the entire observing time, and call that number $N_{coinc}$.

To classify these coincident glitch events, we find the classification reported by GravitySpy for the Omicron event in the strain channel and enforce that the confidence of the classification is greater than 0.9.
We then assume that this classification applies to the Omicron event, and the relevant coincident iDQ event as well.
In addition to known morphological glitch classes, GravitySpy additionally includes one classification called "No Glitch". 
This class is meant to truly classify times without glitches present, but the GravitySpy model used in this work has recently been found to confidently assign this label to times which clearly have excess power in their spectrograms that does not necessarily match any of the other classes.
Three examples of this classification can be seen in Figure \ref{fig:no_glitch_qscan}.
The leftmost panel shows a correct classification of No Glitch.
While there may be some excess power, it is not noticeable above the noise background nor well-localized and is therefore not a glitch.
The middle panel shows a time which is a glitch, but which does not match the morphology of any glitch class known to GravitySpy, and is instead added into the No Glitch category.
In the far right panel, we show a clear Extremely Loud glitch that mis-classified into the No Glitch class.
It's possible that the overlap of the Extremely Loud glitch with loud repeating whistle glitches confused the classifier, but it is certainly not devoid of a glitch.
In order to avoid confusion, we therefore throughout this paper will reference this "No Glitch" category as simply "Unknown" as these times may or may not contain glitches.

Any iDQ event which crosses the log-likelihood chosen, but which does not coincide with an Omicron event with SNR greater than 10, we assume to be identification of a time which does not contain a glitch and is therefore a false alarm.
The iDQ training sets meanwhile use times with an Omicron threshold less than 5.5 or no Omicron event at all, for identification of times which do not contain glitches.
Times which fall in the range of Omicron SNR 5.5 and 10 are not confidently true glitches, but do contain excess power in their spectograms as identified by Omicron.
Thereby assigning false alarms in this work to be any event not coincident with an Omicron event with SNR greater than ten is a conservative estimate.
We then calculate a false alarm rate as the number of false alarms in a given time period over the total detector observing time during that period.

We then additionally calculate the rate at which these coincidences would appear for two Poisson event generators.
If the rate at which coincidences actually appear is greater than the Poisson rate, then we can conclude a true correlation between the events which iDQ and Omicron report with a estimate of the significance as the ratio of the coincident rate to Poisson rate.
We calculate this Poisson rate as follows:
\begin{equation}
    \sigma_{P}(\mathcal{L}) = \sigma_{omic} * \sigma_{iDQ}(\mathcal{L}) * T
\end{equation}
where $\sigma_{omic}$ is the rate of omicron events with SNR greater than 10, $\sigma_{iDQ}(\mathcal{L})$ is the rate of iDQ events above the chosen log-likelihood threshold, $\mathcal{L}$, and T is the total observing time.

We further construct a probability on the data being a glitch given what we have observed using Bayesian statistics:
\begin{equation}
    P(\text{glitch}|\text{data}) = \frac{P(\text{glitch}) P(\text{data}|\text{glitch})}{P(\text{data})}
\end{equation}
Where $P(\text{glitch})$ is the prior and $P(\text{data})$ the normalization.
$P(\text{glitch})$ can be taken as the probability of observing a glitch independent of iDQ and in this study is the probability of any time being flagged by Omicron with an SNR greater than 10.
Assuming glitches are Poisson distributed, and that Omicron is effective at identifying them, we can calculate the probability of observing at least one glitch per coincidence window as:
\begin{equation}
    P(\text{glitch}) = (1 - e^{\sigma_{omic} * w_{coinc}})
\end{equation}
where $\sigma_{omic}$ is the omicron glitch rate rate, and $w_{coinc}$ is the half second coincidence window as described before.
$P(\text{data})$ is then the probability of seeing iDQ data above the threshold we chose, or the probability that the time of interest has been flagged by iDQ.
We can calculate this probability similarly:
\begin{equation}
    P(\text{data}) = (1 - e^{\sigma_{idq}(\mathcal{L}) * w_{idq}})
\end{equation}
where again $\sigma_{idq}(\mathcal{L})$ is the rate of iDQ events above the rank threshold, and $w_{idq}$ is the clustering window as described previously.
Finally, $P(\text{data}|\text{glitch})$ is then the probability of having seen the iDQ data given that a glitch is present, or the probability of the iDQ data occurring given that there is also an Omicron glitch flagged.
In our study, this must be dependent on the ratio of total time covered by coincident events, and the total time covered by glitch events.
In other words:
\begin{align}
    P(\text{data}|\text{glitch}) &= (1 - e^{\eta_{coinc}(\mathcal{L})}) \\
    \text{where } \eta_{coinc}(\mathcal{L}) &= \frac{\sigma_{coinc}(\mathcal{L}) * w_{coinc}}{\sigma_{omic} * w_{coinc}} = \frac{\sigma_{coinc}(\mathcal{L})}{\sigma_{omic}}
\end{align}

The final probability can then be constructed as:
\begin{equation}
\label{eq: p(glitch|data)}
    P(\text{glitch}|\text{data}) = \frac{(1 - e^{\sigma_{omic} * w_{coinc}}) * (1 - e^{\eta_{coinc}(\mathcal{L})})}{(1 - e^{\sigma_{idq}(\mathcal{L}) * w_{idq}})}
\end{equation}
This estimate of the probability has the benefit of being based solely on counting statistics, meaning the underlying distributions can be collected cumulatively for real-time analysis without loss of latency.
This estimate has the downside, however, of being dependent on the coincidence, and clustering windows chosen to estimate the duration of glitches identified by iDQ and Omicron.
As constructed, the maximum $P(\text{glitch}|\text{data})$ obtainable is dependent on the ratio of the coincident and idq clustering windows.
For example, implementing a coincident window half the size of the idq clustering window results in a maximum obtainable $P(\text{glitch}|\text{data})$ of one half.
To mitigate this, the constant windows could instead be replaced by time-based segments flagged by iDQ constructed with the un-clustered iDQ timeseries, and by Omicron constructed with the individual glitch durations.
Then, the time covered by coincident events could be given as the overlap between the two sets.
The implementation for low-latency, and this definition using segment logic, has been left to future work.

\begin{figure*}[htb]
    \centering
    \includegraphics[width=2.\columnwidth]{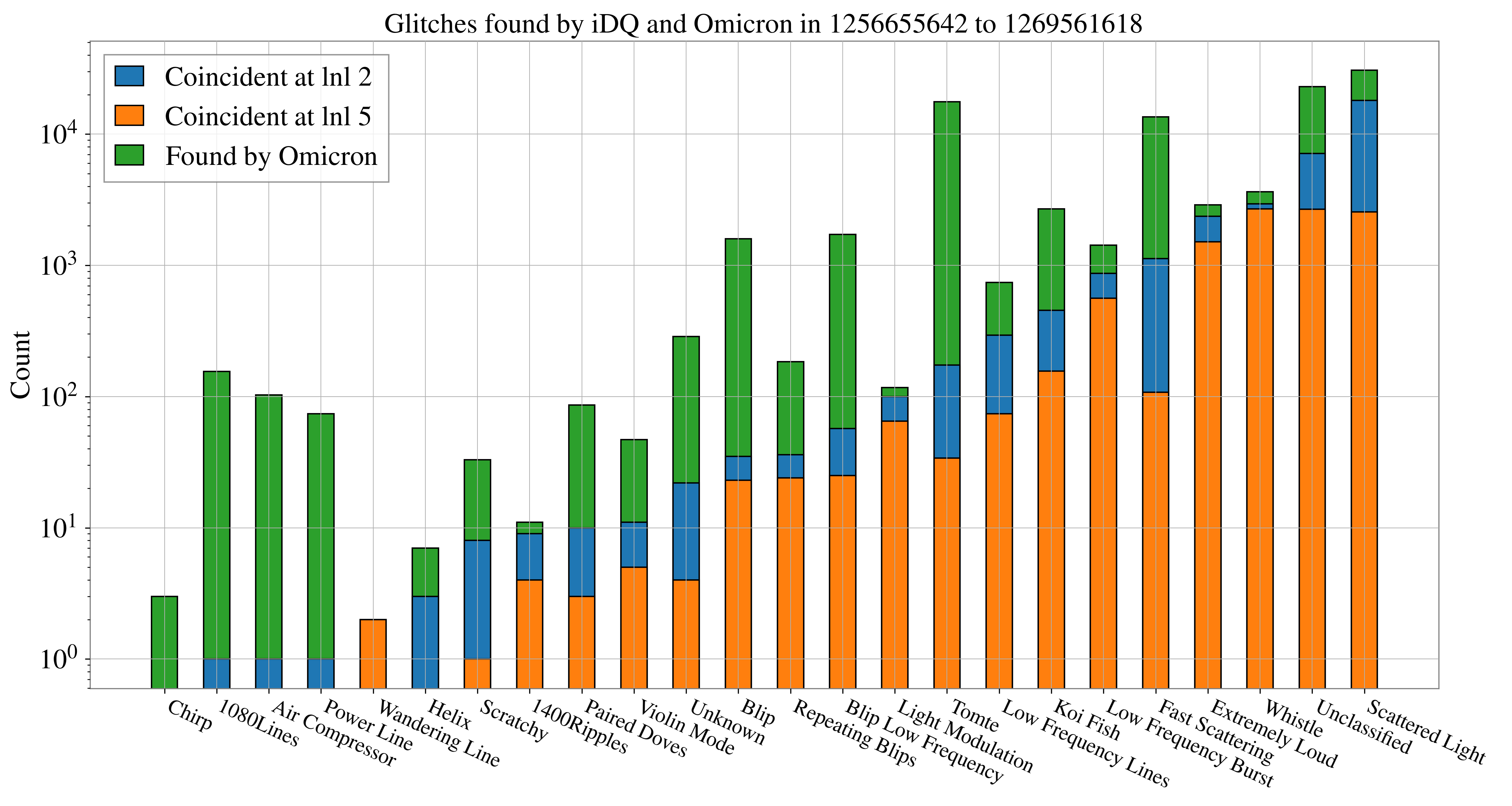}
    \caption{Glitch classification of event times in O3b. In blue and orange, the coincident times between Omicron triggers with snr greater than 10, and the times which pass two different thresholds on iDQ log-likelihood. In green, the times in strain data where Omicron reported SNR greater than 10, but not passing the log-likelihod threshold in iDQ. As shown by the orange and blue bars closely matching the green, there are several glitch classifications which iDQ seems particularly good at identifying including Scattered Light, Whistle, Extremely Loud, and Low Frequency Burst. This means that during O3b, iDQ likely had extremely effective witnesses for these glitch types, while there may not have been good auxiliary witness channels for others, like Tomte glitches.}
    \label{fig:gspy_class_hist}
\end{figure*}

\subsubsection{Results}
We take our results over all of O3b, or November 2019 to March 2020 from the LIGO Livingston detector.
During this time, there were 100,512 number of departures identified by Omicron with a SNR greater than 10 available in the Omicron database.
78,436 of these additionally had available classifications by GravitySpy with a confidence greater than 0.9.
Using the clustering and coincidence methods described in section \ref{sec: glitch_identification_methods}, iDQ identified 39,398 (39 percent) at a log-likelihood threshold of 2, and 11,915 (12 percent) at a threshold of 5.
It is then evident that iDQ identifies only a fraction of the total number glitches identified by Omicron, but this is expected.
iDQ can only identify glitches for which there are auxiliary channels that reliably predict their presence.
If there is no auxiliary channel activity in any of iDQ's witness channels at the time of a glitch, then iDQ can never report on it.
This can be the case for glitch types whose source channels are not currently monitored by an auxiliary channel in the witness list, whose source may only register quietly in the currently monitored channels, or whose source is not monitored at all by any current auxiliary channel.

The classifications of these coincident events can be seen in Figure \ref{fig:gspy_class_hist} in blue and orange while all of the Omicron events are shown in green.
As mentioned previously, the Unclassified category comes from coincident events which did not have a classification from GravitySpy with a confidence of more than 0.9.
During O3b, it is clear from Figure \ref{fig:gspy_class_hist} that Scattered Light glitches were the main category of glitches plaguing the detectors with over 30,000 present while Tomte and Fast Scattering glitches are closely tied for second.
Additionally, this figure shows that iDQ identifies a large fraction of some of the most common glitch types like Scattered Light, Whistle, Extremely Loud, and Low Frequency Burst while it struggles to identify others like Tomte, Fast Scattering and Koi Fish.
As previously mentioned, this likely means that there are auxiliary channels which reliably record the presence of Scattered Light, and Whistle glitches while there may have been a lack of such channels for Tomtes and Blips.

Additionally, the fact that iDQ does not report on a single Chirp glitch at either threshold is, in fact, a feature.
Chirps, as labeled here, are times in the strain channel which have a high SNR and whose morphology mimics a chirp shape.
In other words, these are times which could very well be real gravitational waves as real gravitational wave events also follow the chirp morphology in strain data.
As a data quality product, it is desirable that iDQ does not identify these times.
For example, if the output of iDQ were used to generate vetoes for search pipelines, we would not want times which could be gravitational waves to be included in the vetoed set.
This is an advantage of using software like iDQ which depends only on the auxiliary channels of the detector, and which therefore is insensitive to gravitational waves versus software like Omicron which directly analyzes the strain channel and identifies both real gravitational waves and glitches without delineation.

\begin{figure*}[htb]
    \centering
    \includegraphics[width=2.\columnwidth]{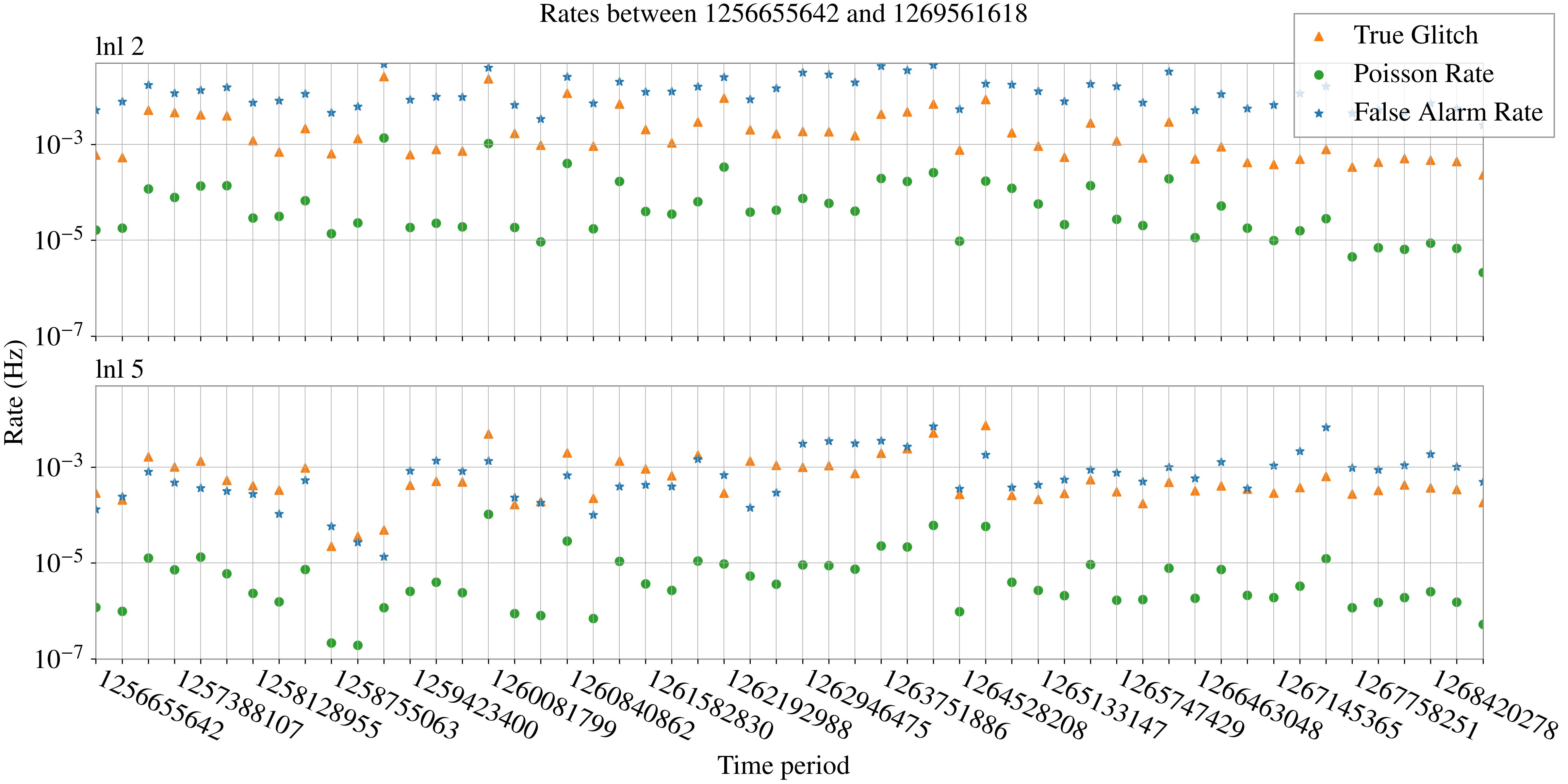}
    \caption{True glitch (orange triangle), false alarm (blue star), and Poisson (green circle) rates as reported by iDQ at a log-likelihood threshold of two (top) and five (bottom) over the course of O3b broken up by approximately five day periods. A true glitch is considered to be an iDQ time crossing the threshold which is coincident with an SNR greater than 10 Omicron time while a false positive is one not coincident with such an event. More details on this delineation are discussion in \ref{sec: glitch_identification}. Notably across time periods, the true glitch, or coincidence, rate is always at least an order of magnitude more than the poission rate implying a true correlation between iDQ and Omicron triggers. Additionally, the false alarm rate is about the same as, or higher than the glitch rate at the log-likelihood threshold of 2. However, at a threshold of 5, the this relationship begins to switch for some time periods.}
    \label{fig:o3b_rates}
\end{figure*}

In Figure \ref{fig:o3b_rates}, we show the false alarm (blue star), Poisson (green circle), and coincident glitch (orange triangle) rates at a log-likelihood threshold of two (top) and five (bottom) as a function of time during the course of O3b.
Although this data covers all of O3b, it was broken into shorter chunks for evaluation, defined by convention within the LIGO Collaboration.
Each chunk of data corresponds to about two weeks of coincident observing time, and three bins were in the offline analysis resulting in evaluation segments, or each x tick, being between 2-4 days apart.
Every three points then correspond to the same chunk of data, and the start of a new chunk is delineated with a GPS time label.
At both thresholds, the coincident glitch rate is always at least one order of magnitude larger than that of the Poisson rate.
This shows that the coincidences formed between iDQ and Omicron are more significant than random chance, and are truly correlated.
As the log-likelihood threshold for iDQ is increased, both the glitch and false alarm rates decrease - a natural result of the increasing confidence a higher log-likelihood corresponds to.
Additionally, as the threshold increases, the true glitch rate is more frequently higher than the false alarm rate than at the lower threshold, again showing the increase in confidence.

At the higher log-likelihood threshold, the variation over time is especially notable.
The difference in time between points is only a couple days, and there are occasional jumps in rate more than an order of magnitude between neighboring points.
This highlights the occasionally extreme variation in the noise background of the detectors even over the course of just a few days and the challenges that detector characterization experts face in characterizing this behavior.
Additionally, the latter half of this plot reveals an interesting behavior change in both the detectors and iDQ.
The glitch rate peaks at both thresholds just before 1264528208, or February 2020.
Just after, however, the variation in rate at the detectors settles significantly and at both thresholds the false alarm rate is always higher than the coincident rate.
This could point to some notable change in the underlying behavior of the detectors, such as the addition reaction chain (RC) tracking, around that time which is propagating into the effectiveness of the auxiliary channels which iDQ uses.

\begin{figure*}[htb]
    \begin{subfigure}
        \centering
        \includegraphics[width=.45\linewidth]{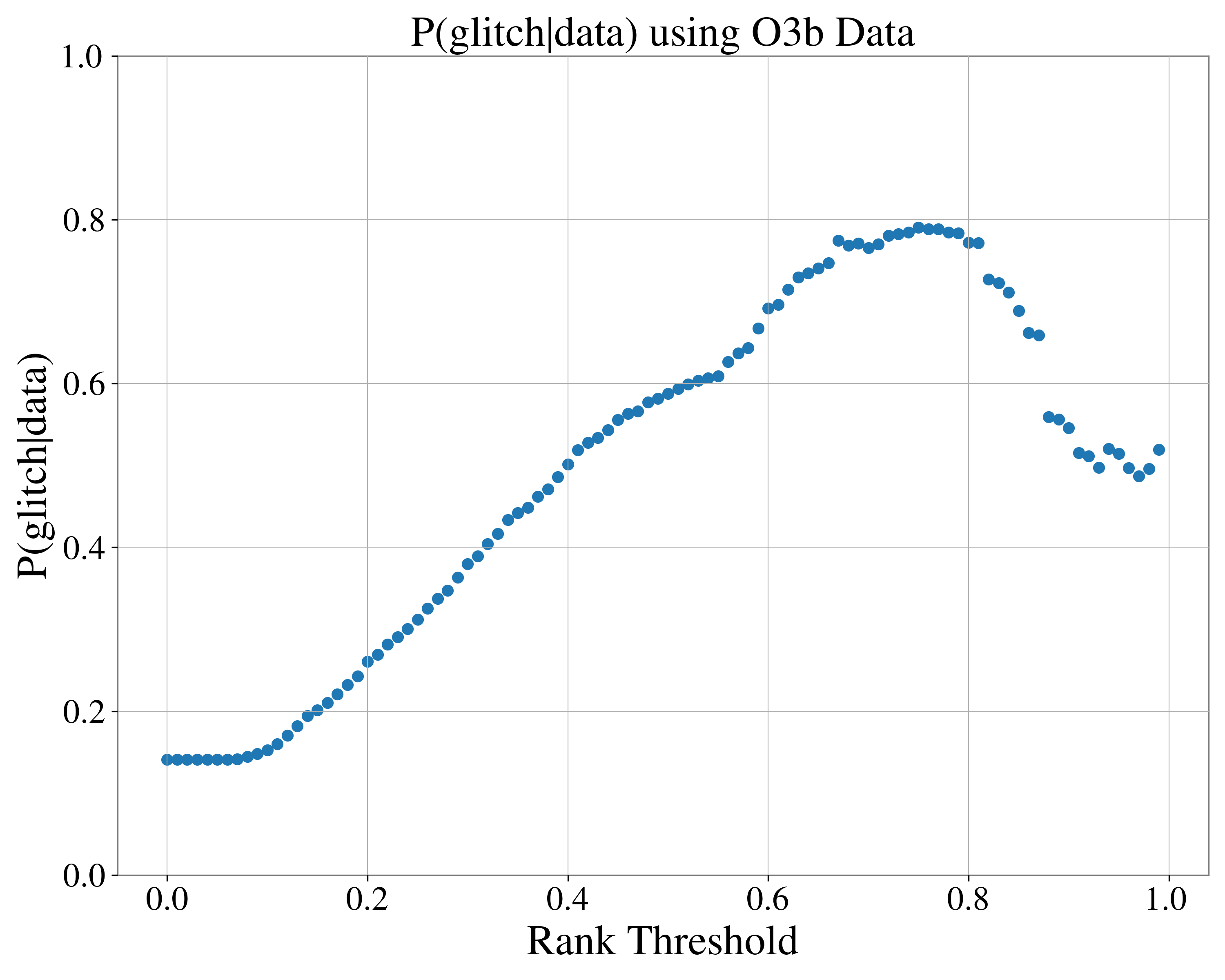}
    \end{subfigure}
        \begin{subfigure}
        \centering
        \includegraphics[width=.45\linewidth]{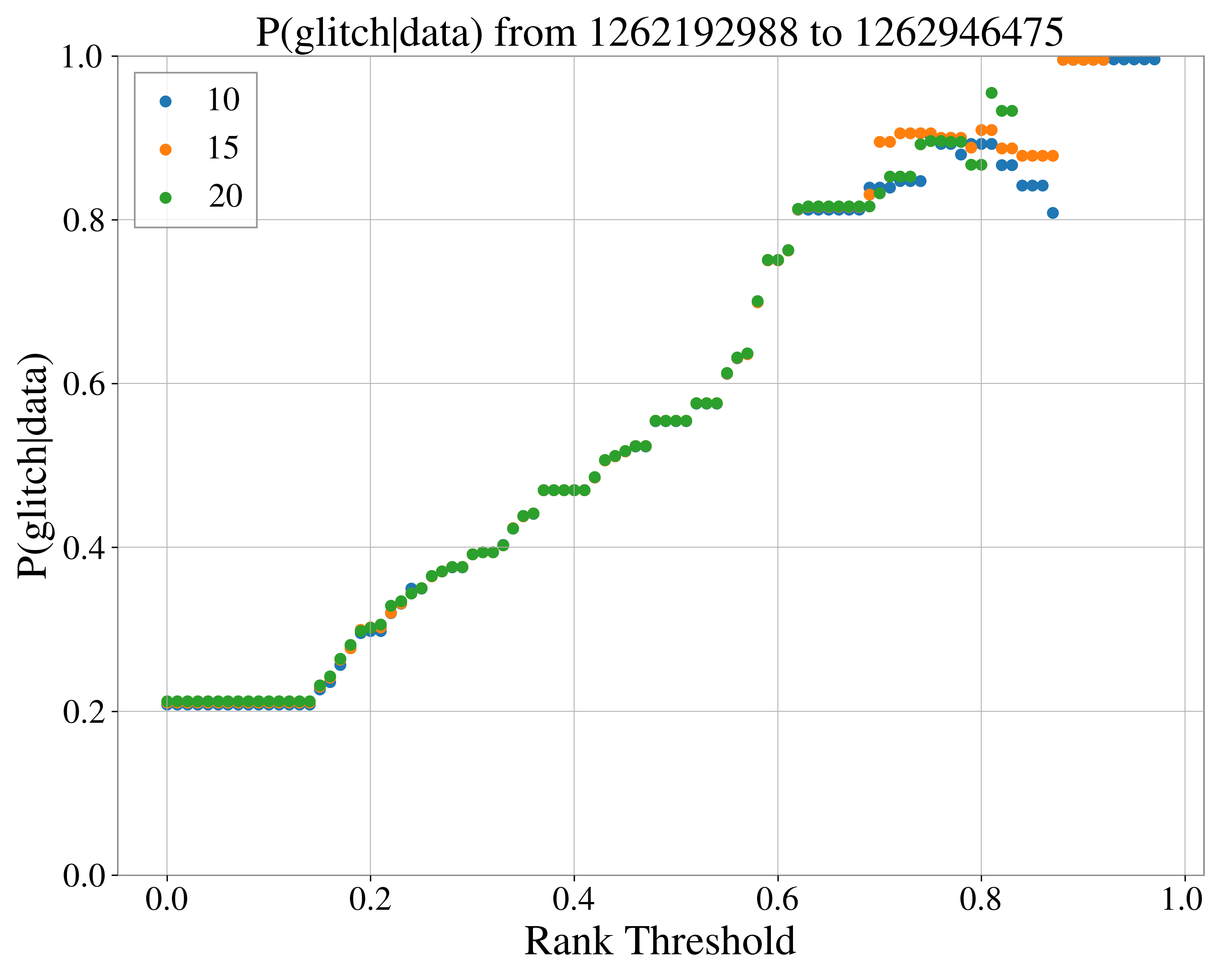}
    \end{subfigure}
    \caption{P(glitch$\mid$data) as defined in equation \ref{eq: p(glitch|data)} using coincident events over the entire course of O3b (left) and one approximately two week period (right) between Omicron with SNR $>$ 10 and iDQ events at a variety of thresholds on rank. On the left, this plot demonstrates the steadily increasing probability of a time being a glitch with increasing iDQ rank up to a certain point around a rank of 0.8. After this, the implementation of use-percentage in the combined rank value causes a number of vetoes with very small Poisson significance to be ranked highly, thereby causing a sharp increase in false alarms at high rank. On the right, a demonstration of how enforcing a minimum Poisson significance on vetoes mitigates this behavior. Already at a threshold of 10 on the Poisson significance, the turning point is mitigated. Then at a threshold of 20, the turning is completely removed.}
    \label{fig:P(glitch|data)}
\end{figure*}

In Figure \ref{fig:P(glitch|data)}, we show the results of Equation \ref{eq: p(glitch|data)} using data from the entire course of O3b on the left and over about a two week period on the right.
The x-axis in both plots represents the varying  threshold applied to the iDQ event times used in the calculation of coincidence and data rates with the combined rank shown in equation \ref{eq:new_rank} as the metric.
The plot shows the steadily increasing probability of a glitch time with increasing rank up to a rank \~0.8, or log-likelihood of about 200.
This is expected as the increasing log-likelihood corresponds to an increasing confidence by iDQ of a glitch presence.
The plateau at the smallest ranks between 0 and about 0.1 are because events with log-likelihood values less than two, or about a rank of 0.1, were not included in this analysis and therefore we see a plateau extending to 0 at the logl 2 value.

After rank 0.8, there is additionally a dip for higher threshold times where we would expect a continuation of the upward trend.
This is because the current implementation of the combined rank with a 2/3 weight on use percentage favors vetoes which activate as few as a single time during the training period which happens to coincide with a glitch time.
This gives the veto a use percentage of 100\% as it correctly flagged a glitch during the single time it was active, and is therefore highly ranked.
However, vetoes such as these do not generally predict glitch presence, and therefore cause false alarms when applied to real data, causing the turn we see at high rank.

To mitigate this, we enforce that vetoes must also have a minimum Poisson significance to be considered.
The image on the right of Figure \ref{fig:P(glitch|data)} shows the results for a single chunk, or about two weeks, out of a total 17 chunks of data from O3b with a variety of Poisson minima enforced.
At the smallest minmum Poisson significance value, we see the highest rank $P(\text{glitch})$ values go to what we would expect, but with those at slightly lower rank still being affected.
As you increase the minimum Poisson significance, however, we see the effect extend to the lower rank values as well until eventually at a minimum of 20, the slope extends fully across all ranks.
While the minimum Poisson significance has not been implemented across the O3b data in this work, this change will not affect the results shown here as the thresholds applied were at low log-likelihood values which live at an equivalent rank much smaller than 0.8.

Compared to the prior probability given by Omicron of 3\% for this same data, the $P(\text{glitch})$ values across the rank space are an improvement over using Omicron information alone.
At the lowest threshold considered in this paper of a log-likelihood of 2, or equivalently rank of around 0.1, $P(\text{glitch})$ already sits at a value of 14\%.
This shows the power in combining results from across both data quality products.

\subsection{Auxiliary Channel Witness Identification}
\label{sec: aux witness}

An additional benefit of an iDQ identification is that we can glean further insight into these glitch types through the auxiliary channels which monitor them.
While GravitySpy allows us to classify times identified by iDQ and Omicron, we further this classification by combining the GravitySpy label with feature information on auxiliary channels used by iDQ.
Each time analyzed by iDQ is assigned a rank via an Ordered Veto List by OVL as described in section \ref{sec:background}.
In this analysis, we take the model applied at a time of interest and then look at all of the vetoes which pass the log-likelihood threshold in the relevant Ordered Veto List to discover which correlated auxiliary channels were active.

The channels associated with the vetoes which pass the log-likelihood threshold are then by definition those which make the most effective vetoes, or those whose activity often corresponds to excess power in the strain channel.
It is not unreasonable then to assume that the channels which made the best vetoes for a certain glitch class at the very least frequently record instances of it, and at best could monitor the subsystem which is a possible source of that glitch class.
Under this assumption, we look across all times of interest and count how often individual channels are active during a particular class of glitch.

\begin{figure}
    \centering
    \includegraphics[width=.45\textwidth]{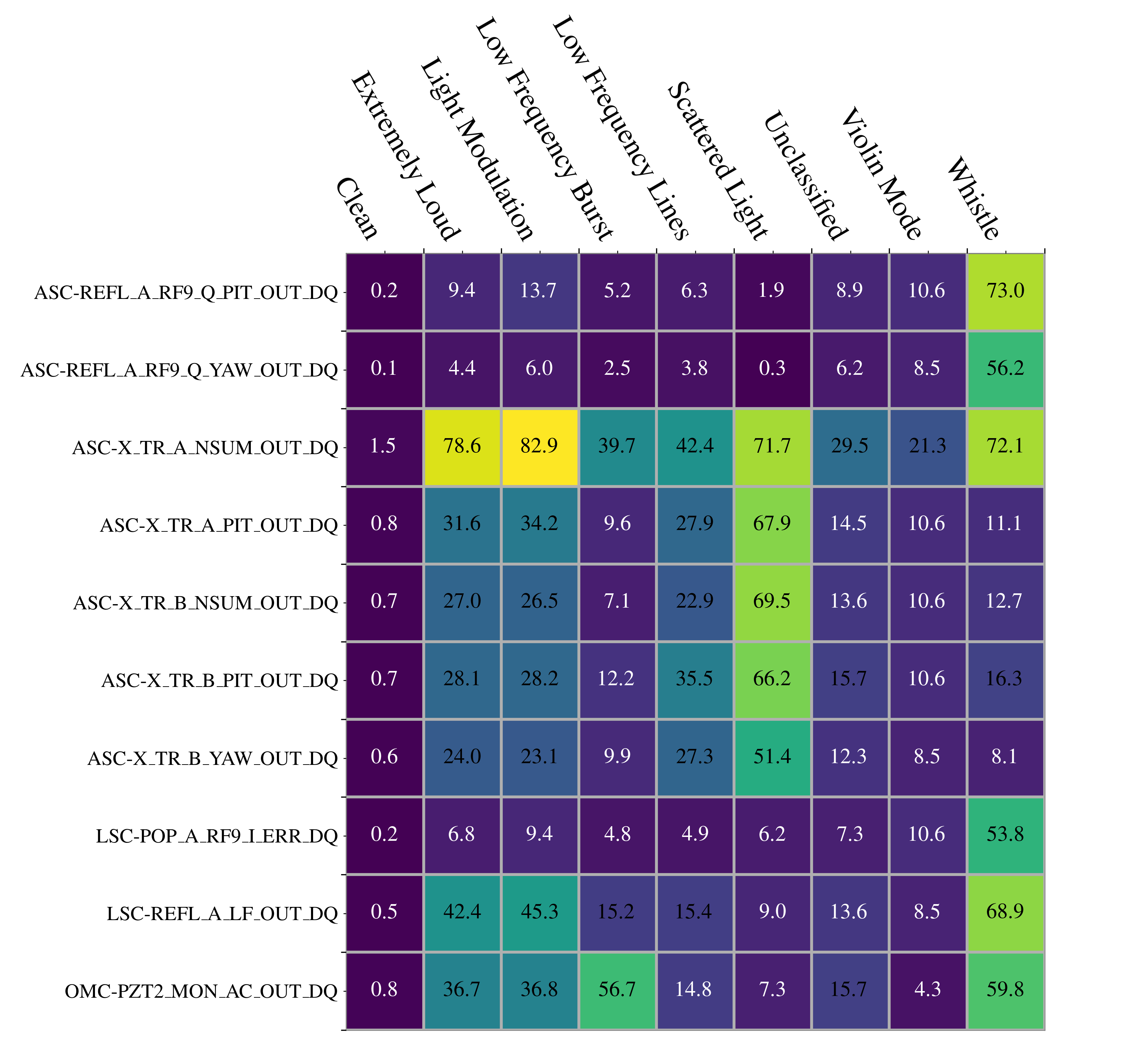}
    \caption{Auxiliary channels that form vetoes in the associated OVL model at least 20\% of the time at a log-likelihood of 2 or more for a variety of glitch classes.
    The included glitch classes are caught in coincidence by iDQ and Omicron at least ten times over the course of O3b with an iDQ log likelihood of at least 2 and an Omicron SNR of 10 or greater. Additionally included are the auxiliary channel results for the false alarms flagged by iDQ weighted by the total time not covered by Omicron glitches for comparison. Note that the majority of channels which are present in glitch veto lists, are not frequently present in clean veto lists.}
    \label{fig:aux_grid}
\end{figure}

\begin{figure*}
    \centering
    \includegraphics[width=2.\columnwidth]{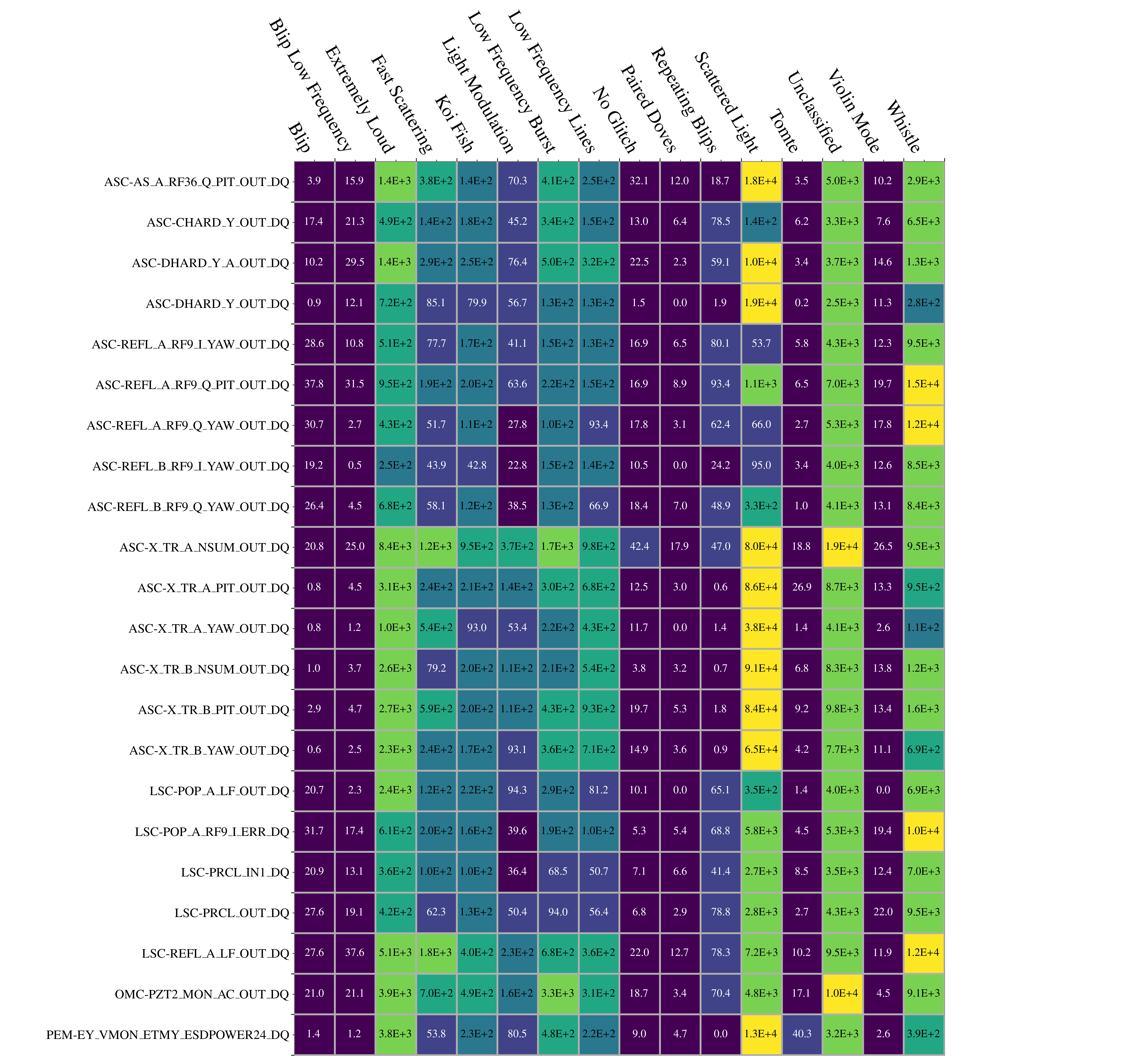}
    \caption{Auxiliary channels which appear in the top 10 vetoes of the associated OVL model with a Poisson significance of at least $5\mathrm{E}{+3}$ for a variety of glitch classes which are caught in coincidence by iDQ and Omicron at least ten times over the course of O3b with an iDQ log likelihood of at least 2 and an Omicron SNR of 10 or greater. Additionally included are the auxiliary channel results from a random set of clean samples for comparison. Even at the higher log likelihood threshold cut-off, the majority of channels present witness some combination of glitch types, but mostly exclude clean samples. A sign that these channels are truly good witnesses of glitches, and not just frequently active.}
    \label{fig:aux_grid_false_alarm}
\end{figure*}

In Figure \ref{fig:aux_grid}, we show this correlation using a log-likelihood threshold of two and weighting by the total number of glitches recorded by Omicron.
In Figure \ref{fig:aux_grid_false_alarm}, we show this same correlation at the same threshold, but instead report the Poisson significance\footnote{Poisson significance is the negative log of the total probability of observing as many or more coincidences between two series of random occurrences than were actually observed. In the context of Figure \ref{fig:aux_grid_false_alarm}, one can assume the total number of times an aux channel triggers and the number of times omicron triggers for a given glitch type are both represented by Poisson processes. The larger the significance the lower the probability of the triggers being coincident by pure chance. Readers are referred to \cite{Essick_2013} for further discussion on Poisson significance. } between channels and a given glitch type.
The former gives an overview of channel performance relative to all glitches of that class and identifies channels which are particularly good witnesses of most glitches in that class.
It is possible, however, that there are multiple sources for the same class of glitch, so the latter plot focuses only on the sources which iDQ identifies instead of on the general performance of these channels.
Additionally, we choose the lower log-likelihood threshold of two in this case to use as much information available to us as possible, and include information on clean samples as a protection against the lower confidence.

On the y-axis of both plots are the channels which are active at least 50\% of the time for at least one glitch class.
On the x-axis are the glitch classes with at least ten coincident events, and which have at least one channel that appears at least 20\% of the time for at least one glitch class.
In Figure \ref{fig:aux_grid} we additionally include one column which reports on the auxiliary channel presence during the false alarms recorded by iDQ and weighted by the total time not covered by coincident glitches e.g. clean time, as there is by definition not coincident Omicron triggers available for False Alarms.
This is a stand-in for the probability that the channel was active during a random non-glitch time.

The Clean column is used to exclude the possibility that a channel seems to be a good witness for glitches simply because it is almost always active.
As described in Section \ref{sec:background}, channels which are active the majority of the time generally get down-ranked by the OVL ranking scheme because the high activity introduces a large deadtime and low use-percentage to the veto.
Therefore, we wouldn't expect these kinds of channels to appear highly ranked in the veto lists and in these plots, but we consider all vetoes passing the log-likelihood threshold in the OVL list without weighting them.
Therefore channels could be frequently ranked middling to low across glitch types, but appear in these plots to have the same significance as one which constantly appears with the highest rank thereby inflating its apparent significance.
The inclusion of auxiliary witnesses for clean  samples is a sanity check on this possibility.
If a channel is frequently present across glitch types, and during clean then we know that its possible this channel is just generally active.
However, if it is present during glitch times and not during clean ones then we can be confident that it a true witness for glitches.

The auxiliary channel ASC-X\_TR\_A\_NSUM\_OUT\_DQ, for example, seems to be one of these extremely active channels.
In both Figure \ref{fig:aux_grid} and \ref{fig:aux_grid_false_alarm}, this channel, which is described below, appears incredibly frequently across glitch types.
This a prime example of a generally active channel which results in a veto with a high efficiency, but middling use percentage. In other words, it creates an efficient veto which flags many glitch classes, but also can flag false alarms and we're seeing that trade-off here.
This kind of veto may appear at middling to low rank in the ordered veto lists, but still cross the log-likelihood threshold of two and therefore appear in this analysis.

The three channels below that one in Figure \ref{fig:aux_grid}, however, are good examples of the opposite.
ASC-X\_TR\_A\_PIT\_OUT\_DQ, ASC-X\_TR\_B\_NSUM\_OUT\_DQ, and ASC-X\_TR\_B\_PIT\_OUT\_DQ each appear over 50\% of the time for all Scattered Light glitches, or with considerable Poisson significances for all coincident Scattered Light glitches over the course of O3b in Fig \ref{fig:aux_grid_false_alarm} while appearing significantly less frequently for all other glitch types.
This implies that while these auxiliary channels may be excellent witnesses of one source of Scattered Light glitches there are likely other sources which do not have good auxiliary witnesses contributing to the overall number flagged by Omicron.
These three channels specifically monitor different pieces of the same subsystem in the Livingston detector.
They are each part of the angular sensing and control subsystem (ASC) in the direction of the X arm, particularly monitoring the transmitted light (TR) on two different photodiodes (A/B).
It follows that channels detecting transmitted light on one of the major axes of the detector would observe a large fraction of scattered light from a variety of sources, including sources from the cavity before the test mass, and systems after the test mass, both orientations from which Scattered Light glitches are known to originate.
Its interesting though that while one channel in this susbsystem set, ASC-X\_TR\_A\_NSUM\_OUT\_DQ, is active across glitch classes, three others catch more exclusively true Scattered Light glitches and this demonstrates the power of this method.
From this information, its safe to say that X arm beam in particular at Livingston during O3b was a source of one of the most common glitches plaguing the detector at the time, particularly the three origins of the specific channels mentioned above.
In fact, this is known to be the case as discussed in detail \cite{LIGO:2021ppb}.
Halfway through O3, ground motion was found to cause variation in the difference between mirrors in the arms, and therefore cause scattered-light to rejoin the main beam registering as Scattered Light glitches.
During the commissioning break between O3a and O3b, maintenance was performed to mitigate this issue which drastically decreased the total number of Scattered Light glitches recorded during O3b comparatively to O3a, but as is clear from this data, some glitches persisted from this source.

ASC-REFL\_A\_RF9\_Q\_PIT\_OUT\_DQ is another excellent witness, particularly of Whistle glitches.
This channel appears in over 70\% of all Whistle glitches recovered during O3b, and has a Poisson significance of $1.5\mathrm{E}{+4}$ in regards to the glitches recovered in coincidence.
Its sister channel ASC-REFL\_A\_RF9\_Q\_YAW\_OUT\_DQ  is also highly active at over 55\% of all Whistle glitches and has a Poisson significance of $1.2\mathrm{E}{+4}$ of the coincident ones.
Both channels again monitor the angular sensing and control grouping, but these wavefront sensors particularly monitor one component of the beat of the sidebands and carrier modes with respect to one another in the reflected direction of the power recycling cavity.
It is interesting that these Whistle glitches are be observed so well by a channel monitoring the reflected direction, as this direction is more often attributed to various kinds of scattering glitches.
It's possible that this channel just happens to be downstream of the system which is actually generating the glitches, and so it is being monitored here despite not being the origin.
However, the very similar monitoring channels on wavefront sensor B show extremely similar correlations to the Whistle glitch types as these two channels, so it becomes even more convincing that a Whistle glitch source lies either in or upstream of these reflected wavefront monitors.

These four channels are additionally active for a handful of other glitch classes as well such as the Blip and Repeating Blip class, although with a weaker correlation.
Blip glitches could have the same sources as Repeating Blips as arguably one is just a more frequent version of the other.
It would be new information though to also include Whistle glitches in that mix.
It is possible that this could be a false correlation if the GravitySpy model, for example, frequently misclassified Whistle glitches as Blips and vice versa, but there is no evidence that that is the case.
Instead, this could be the result of a couple different configurations.
This channel could monitor the physical sources of multiple glitch classes, meaning it could be downstream from other parts of the detector system which individually generate these glitch types.
It could also be that there is one source which this channel monitors that creates many different kinds of glitches.
Generally, we assume that glitch classes are generated by distinct sources, but if this latter situation is true it could hint at correlations between glitch classes previously unconsidered.
With this information alone, it is hard to tell which may be the true case, but either way it demonstrates the usefulness of this method in further characterizing detector behavior.

\section{Conclusion}
In this work, we've discussed improvements to the iDQ batch pipeline and we've demonstrated iDQ's ability to not only identify glitches in strain based solely on auxiliary channel behavior, but also shown its usefulness in identifying auxiliary channels which frequently report on the presence of glitches.

During O3b, it has been shown that iDQ had particularly powerful witness channels Scattered Light, Whistle, and Extremely Loud glitches.
The correlation of these recovered events with the events that Omicron finds has been shown to be frequently two orders of magnitude greater than random chance, confirming that Omicron and iDQ are truly recovering the same events.
Additionally, by analyzing the auxiliary channels alone, iDQ does not identify chirps, or likely real gravitational waves, to be glitches which Omicron reports in the same manner as any other glitch class.
This not only proves the effectiveness of iDQ's identification scheme, but also the worth of its results alongside other glitch identification tools.

We have also introduced a method for calculating the probability that any time is a glitch based on iDQ data.
We have seen that using the current method, the probability peaks at about 70\%, or three orders of magnitude above probability given by Omicron identification alone.
This demonstrates the power of combining results across multiple glitch identification tools, and could be a useful measure of data quality for inclusion in gravitational wave detection pipelines. 
The authors hope to implement this method in the near-future for use in real-time analyses.

We have demonstrated the effectiveness in examining the correlations identified by iDQ between auxiliary channel activity and certain glitch classes.
These correlations hint at possible sources of these glitch types, and when combined with follow-up from commissioners, could be a powerful tool in tracking down the origins of some glitch classes.
Unfortunately, there is no way to know for sure whether any unique channel points to a true origin or whether a subsystem somewhere else in the detector is causing a glitch which then may propagate until witnessed by an unrelated auxiliary channel.
Either way, channels like the ones we have mentioned could point to possible starting points for commissioners and detector characterization experts to begin looking for the sources of these extremely common glitches.
If this analysis had been performed during an observing run, follow-up could have been done by commissioners on these channels, and these glitches could have been potentially mitigated during regular maintenance, or even during a longer commissioning break.
The authors hope to make the auxiliary channel information available in low-latency in order to potentially impact maintenance and commissioning on the detectors moving forward.

As the detectors, glitch types, and auxiliary channels evolve between observing runs, iDQ will evolve with them, and the need for robust data quality information will only grow and sensitivity of the detectors increase.
While this work demonstrates iDQ's past performance, it is heavily reliant on the quality of its auxiliary witnesses and as these change, so too will the types of glitches iDQ can identify and the efficiency at which it does so.
Already, the LIGO Scientific and Virgo Collaborations have begun a fourth observing run in which the group has made many changes to the detectors and have a planned commissioning break before the second half in which more changes will be made.
While results from the current observing run will of course vary the ones shown, iDQ has proven to be a reliable pipeline providing both probabilistic glitch identification, as well as glitch source identification, and it will continue to do so throughout O4 and beyond.

\begin{acknowledgments}
The authors thank Gabriela Gonzalez for valuable comments on the draft of this work as well as Derek Davis, Chad Hanna, Siddharth Soni, and Aaron Zimmerman for helpful discussions.
M.T. acknowledges support from the U.S. National Science Foundation through grants PHY-2309085 and PHY-2409448. Z.Y. acknowledges support through NSF grant NSF-PHY-2110509. The authors are grateful for computational resources provided by the LIGO Laboratory and supported by National Science Foundation Grants PHY-0757058 and PHY-0823459. This material is based upon work supported by NSF’s LIGO Laboratory which is a major facility fully funded by the National Science Foundation. LIGO was constructed by the California Institute of Technology and Massachusetts Institute of Technology with funding from the National Science Foundation (NSF) and operates under cooperative agreement PHY-1764464. The authors are grateful for computational resources provided by the Pennsylvania State University’s Institute for Computational and Data Sciences (ICDS) and the University of Wisconsin Milwaukee Nemo and support by NSF PHY-2011865, NSF OAC2103662, NSF PHY-1626190, NSF PHY-1700765, and NSF PHY-2207728. This paper carries LIGO Document Number LIGO-P2400455.

\end{acknowledgments}

\bibliographystyle{apsrev4-2}
\bibliography{main.bbl}

\end{document}

%% file: authors.tex
\author{Rachael Huxford}\thanks{These authors contributed equally}

\affiliation{Department of Physics, The Pennsylvania State University, University Park, PA 16802, USA}
\affiliation{Institute for Gravitation and the Cosmos, The Pennsylvania State University, University Park, PA 16802, USA}

\author{Richard N. George\textsuperscript{†}}\thanks{These authors contributed equally}

\affiliation{Center for Gravitational Physics, University of Texas at Austin, Austin, TX 78712, USA}

\author{Max Trevor}\thanks{These authors contributed equally}

\affiliation{Department of Physics, University of Maryland
John S. Toll Physics Building, College Park, MD 20742, United States}

\author{Zach Yarbrough}
\affiliation{Department of Physics, Louisiana State University, 202 Nicholson Hall Baton
Rouge, LA 70803 USA}

\author{Patrick Godwin}
\affiliation{LIGO Laboratory, California Institute of Technology, MS 100-36, Pasadena, California 91125, USA}